\newif\ifsubmode
\submodefalse

\newif\ifprintfig
\printfigtrue

\newif\ifemulate
\emulatefalse

\ifsubmode
  \documentclass[12pt,preprint]{aastex}
  \received{}
  \accepted{}
  \journalid{}{}
  \articleid{}{}
\else
   \documentclass{emulateapj}
   \submitted{{\it Accepted for publication in ApJ}}
\fi

\renewcommand\email\texttt

\def\spose#1{\hbox to 0pt{#1\hss}}
\def\lta{\mathrel{\spose{\lower 3pt\hbox{$\sim$}}
    \raise 2.0pt\hbox{$<$}}}
\def\gta{\mathrel{\spose{\lower 3pt\hbox{$\sim$}}
    \raise 2.0pt\hbox{$>$}}}

\begin{document}

\slugcomment{\sc Accepted to \it Astrophysical Journal}

\shorttitle{\sc Bo\"otes II} \shortauthors{Walsh et al.}

\title{Bo\"otes II ReBo\"oted: An MMT/MegaCam Study of An Ultra-Faint Milky Way Satellite}

\author{S.\ M. Walsh\altaffilmark{1}, B. Willman\altaffilmark{2,3},
  D. Sand\altaffilmark{4,5}, J. Harris\altaffilmark{4},
  A. Seth\altaffilmark{2}, D. Zaritsky\altaffilmark{4},
  H. Jerjen\altaffilmark{1}}

\altaffiltext{1}{Research School of Astronomy and Astrophysics,
  Australian National University, Cotter Road, Weston, ACT 2611,
  swalsh@mso.anu.edu.au}

\altaffiltext{2}{Harvard-Smithsonian
Center for Astrophysics, 60 Garden Street Cambridge, MA 02138}
\altaffiltext{3}{Clay Fellow, beth.willman@gmail.com} 

\altaffiltext{4}{Steward Observatory,
University of Arizona, 933 North Cherry Avenue, Tucson, AZ, 85721}

\altaffiltext{5}{Chandra Fellow}

%\newpage

\begin{abstract}

  We present MMT/Megacam imaging in Sloan $g$ and $r$ of the extremely
  low luminosity Bo\"otes II Milky Way companion. We use a bootstrap
  approach to perform robust measurements of, and uncertainties on,
  Bo\"otes II's distance, luminosity, size, and morphology.
  Comparisons with theoretical isochrones and empirical globular
  cluster fiducials show that Bo\"otes II's stellar population is old
  and metal-poor ([Fe/H] $\lta$ -2).  Assuming a stellar population
  like that of M92, Bo\"otes II is at a distance of 42 $\pm$ 2 kpc,
  closer than the initial published estimate of 60 $\pm$ 10 kpc.  This
  distance revision, combined with a more robust measurement of
  Bo\"otes II's structure with a Plummer model (exponential model)
  results in a more compact inferred physical half-light size of
  $r_h\simeq 36 (33) \pm 9 (10)$\,pc and lower inferred luminosity of
  $M_V\simeq-2.4 (-2.2) \pm 0.7 (0.7)$ mag.  The revised size and
  luminosity we calculate move Bo\"otes II into a region of
  size-luminosity space not previously known to be occupied by old
  stellar populations, but also occupied by the recently discovered
  Milky Way satellites Willman 1 and SEGUE 1. We show that the
  apparently distorted morphology of Bo\"otes II is not statistically
  significant given the present data.  We use a tidal argument to
  support a scenario where Bo\"otes II is a dwarf galaxy (dark matter
  dominated) rather than a globular cluster (not dark matter
  dominated), although the uncertainty on the $M/L$ we infer for
  Bo\"otes II is substantial. Moreover, we can not rule out that
  Bo\"otes II is a star cluster on the verge of disruption, such as
  Palomar 5.

\end{abstract}


%\newpage

\section{Introduction}

Over the last 5 years, the Sloan Digital Sky Survey has been
extensively searched for extremely low surface brightness dwarf
spheroidal galaxies. These searches use the catalog of stellar sources
to identify spatial overdensities of the old, metal-poor stars
characteristic of these dwarfs. To date, these searches
\citep[e.g][]{Willman02,koplf} have resulted in the discoveries of
fourteen new Milky Way satellites.

Most of these new objects have total luminosities less than the median
luminosity of the Milky Way's globular clusters ($M_V \sim -7$), but
have sizes characteristic of known dwarf spheroidals ($r_{half}
\gtrsim$ 100 pc), complicating their classifications as either star
clusters or dwarf galaxies.  In this paper, we assert that the
physical distinction between a globular cluster and a dwarf
galaxy is that a dwarf galaxy is, or was at some point, the primary
baryonic component of a dark matter halo whereas a globular cluster
was not.  Nine of the 14 new satellites were originally classified as
dwarf spheroidals because they had scale sizes $\gtrsim$ 100 pc
(Bo\"otes, Canes Venatici, Canes Venatici II, Coma Berenices,
Hercules, Leo IV, Leo T, Ursa Major and Ursa Major II;
\citealp{bootes,cvn,sakamoto06,quintet,uma,uma2}).  Follow-up spectroscopic
studies demonstrated that they indeed appear to require dark matter to
explain their kinematics, securing their classification as dwarf
galaxies \citep{simon07,spectro,strigari07,walker07}. Koposov 1 and 2
\citep{k12} have scale sizes of only 3 pc, and have thus been
classified as new globular clusters.

The classification of the other three new satellites, Segue 1,
Willman 1, and Bo\"otes II \citep[$M_V \sim
-2.5$;][]{quintet,willman1,boo2}, has been less straightforward.
They are old stellar populations with sizes intermediate
between known Milky Way globular clusters and dwarf spheroidals, but have
fewer stars than nearly any known galaxy or globular cluster.
Although initial estimates based purely on SDSS data placed Bo\"otes
II close to Coma Berenices in size-luminosity space at $(\log
(r_h/pc),M_V)=(1.85,-3.1)$, in this paper we present estimates based
on a more robust algorithm and on deeper, MMT/MegaCam imaging in $g$
and $r$ that shift Bo\"otes II's size and luminosity closer those of
Willman 1 and Segue 1.

Despite its tiny luminosity, spectroscopic [Fe/H] estimates and
kinematic studies of Willman 1 have provided support that it may
require dark matter to explain its properties, thus classifying it as
a dwarf galaxy (\citealt{spectro,strigari07b}). Whether or not these
three Milky Way companions with only $\sim$ 1000$L_{\odot}$ are
galaxies or globular clusters is of fundamental import to both our
understanding of galaxy formation at the smallest scales and to our
understanding of the size and mass scale of dark matter clustering.
Although the extent to which tides have effected Willman 1's
present-day luminosity is uncertain, this object raises the questions:
Are we for the first time seeing the low luminosity and mass limit of
galaxy formation?  If so, what properties can we infer for the dark
matter halos that host such galaxies?  Segue I and Bo\"otes II
presently lack their own published spectroscopic studies with which to
evaluate these scenarios.  However, rigorous derivations of the
detection limits of the most recent SDSS searches for such objects
(\citealt{koplf}; Walsh, Willman \& Jerjen {\it in prep}) show that
many similar objects could remain yet undiscovered around the Galaxy,
underscoring the importance of understanding their physical
properties.

With the study presented in this paper, we aim to provide the first
robust measurements of the basic properties (distance, luminosity,
structure) of Bo\"otes II and its stellar population, and to evaluate
the present evidence for its classification.  In \S2, we describe
MMT/MegaCam observations of Bo\"otes II, data reduction, and
artificial star tests.  In \S3, we use these data to derive revised
estimates of Bo\"otes II's properties, to verify that its stellar
population is old and metal-poor, and to investigate whether it has a
distorted morphology. We discuss these results and evaluate evidence
for a dwarf galaxy versus globular cluster classification of Bo\"otes II
in \S4.

\section{Data}
\renewcommand{\thefootnote}{\fnsymbol{footnote}} 

We observed Bo\"otes II on June 05 2007 with MegaCam \citep{mcleod00}
on the MMT.  These data were obtained as part of a larger survey
program to image with MMT/Megacam ultra-faint Milky Way satellites.
MMT/MegaCam has 36 chips with 2048x4608 pixels of 0.08$''$/pixel, for
a total field-of-view (FOV) of 24$'$. We obtained 5 180s dithered exposures
in Sloan $g$, and 5 240s dithered exposures in Sloan $r$ in grey
conditions with 1.0 -- 1.2$"$ image quality in the $g$ and 0.9 --
1.0$''$ image quality in the $r$ images.  We reduced the data based on
the method described in Matt Ashby's Megacam Reduction
Guide\footnote[1]{http://www.cfa.harvard.edu/$\sim$mashby/megacam/megacam$\_$frames.html}. Our
reduction relied in part on software written specifically for
MMT/MegaCam data reduction by Brian McLeod.  We used the Sloan Digital
Sky Survey Data Release 6 (SDSS DR6; \citealt{york00,dr6}) stellar
catalog to derive precise astrometric solutions for each science
exposure. We also used the SDSS catalog to derive an illumination
correction in $g$ and $r$ to divide out the variation in zero-point
across MegaCam's FOV.  We use local copies of the SDSS
dataset, maintained at the Harvard-Smithsonian Center for
Astrophysics.

We did a weighted co-addition of the reduced images using
SWARP\footnote[2]{http://terapix.iap.fr/soft/swarp} and then used the
DAOPHOTII/Allstar package \citep{stetson94} to do point source
photometry on the resulting images.  We visually verified the
integrity of the shape and full-width half-max of the PSFs in the
stacked images across the 24$'$ FOV.  Photometry was carried
out using a method similar to that of \citet{harris07}, with the
exception that we used the command-line versions of DAOPHOT and
Allstar rather than the IRAF versions.

To derive the photometric calibration for our data, we first matched
the SDSS stellar catalog to the Allstar catalog for these new
observations. We used the 91 SDSS stars within our field-of-view (FOV)
with $18 < r < 21$ and $0.1 < g-r < 0.8$ to perform the photometric
calibration.  We limited the calibration to stars fainter than $r$ =
18 mag to avoid the saturation limit of the MegaCam data.  We limited
the calibration to stars with colors between $0.1 < g-r < 0.8$ because
the Bo\"otes II member stars resolved in this study (with the
exception of a few possible blue horizontal branch stars) all have
$0.1 < g-r < 0.6$.  There were insufficient SDSS stars in our FOV bluer
than 0.2 mag to determine whether our derived calibration is
appropriate for very blue stars.

We then did a linear least-squares fit for the zero-points and
color-terms, including uncertainties in color and magnitudes on each
star and throwing out 3 sigma outliers.  
\begin{equation}
g = g_{instr} + 7.27 (\pm 0.029) + 0.091 (\pm 0.068) \times (g - r)
\end{equation}

\begin{equation}
r = r_{instr} + 7.33 (\pm 0.025) + 0.074 (\pm 0.054) \times (g - r)
\end{equation}
Uncertainties were derived from a 1000 iteration bootstrap of the
data.  In addition, there is uncertainty in the SDSS zero-points
themselves of about 0.01 mag
\citep{padmanabhan08}. 

Throughout this paper, we adopt SDSS photometry, rather than MegaCam
photometry, for stars brighter than $r = 18.0$ mag.  All magnitudes in
this paper have been extinction corrected with the values from the
\citet{schlegel98} dust maps provided in the SDSS catalog; the median
E($g-r$) along the line-of-sight to Bo\"otes II is 0.02 mag.

We use artificial star tests to measure the photometric errors and
completeness as a function of position in the $g-r$ color-magnitude
diagram (CMD).  Artificial stars are constructed from the $g$ and $r$
point spread functions (PSFs) measured during the data-reduction
process, and are injected into the co-added $g$ and $r$ images using a
uniform grid with spacing in X and Y equal to ten times the full-width
half-max (FWHM) of the PSF (so that artificial stars overlap only
beyond their $10\sigma$ radii).  This fixed geometry imposes a limit
on the number of artificial stars that can be added to the image of
about 18,500.  To build up our number statistics, we inject artificial
stars into twenty copies of the $g$ and $r$ images, randomly
offsetting the grid's zero-point position in X and Y for each
iteration.  This results in a total sample of 370,000 artificial
stars. The $r$ photometry of the artificial stars is drawn randomly
from $\sim$ 18 to 28 mag, with an exponentially increasing probability
toward fainter magnitudes.  To properly characterize the tail of the
completeness function and the impact of blends on the photometric
errors of faint objects, we simulate stars up to three magnitudes
fainter than the nominal faint limit. The $g-r$ color is then drawn
randomly over the range -0.5 to 1.5 mag to determine the $g$
magnitude. We photometer the artificial-star images with the same
photometry pipeline as we used on the science frames.

If an input artificial star is not present in both the $g$ and $r$
Allstar files, then it is flagged as a non-detection for calculation
of the completeness rate.  We applied a strict cut of DAOPHOT
sharpness parameter of $-1 < sharp < 1$ for a star to be included our
analysis, both for the actual Boo II data and for the artificial star
tests. Although this strict sharpness cut yields completeness limits
that are brighter than if we use no cut at all, we found it necessary
to eliminate many galaxy interlopers and provide an improved
measurement of Bo\"otes II's stellar population while not sacrificing
much precision in our quantitative results.  At a $g-r$ color of 0.25
mag, the 50\% and 90\% completeness limits of these data are $r$ =
23.5 and 22.9 mag.

\begin{figure*}[t!]
%\epsscale{0.6}
%\plotone{BooIICMD.paper.eps} 
\plotone{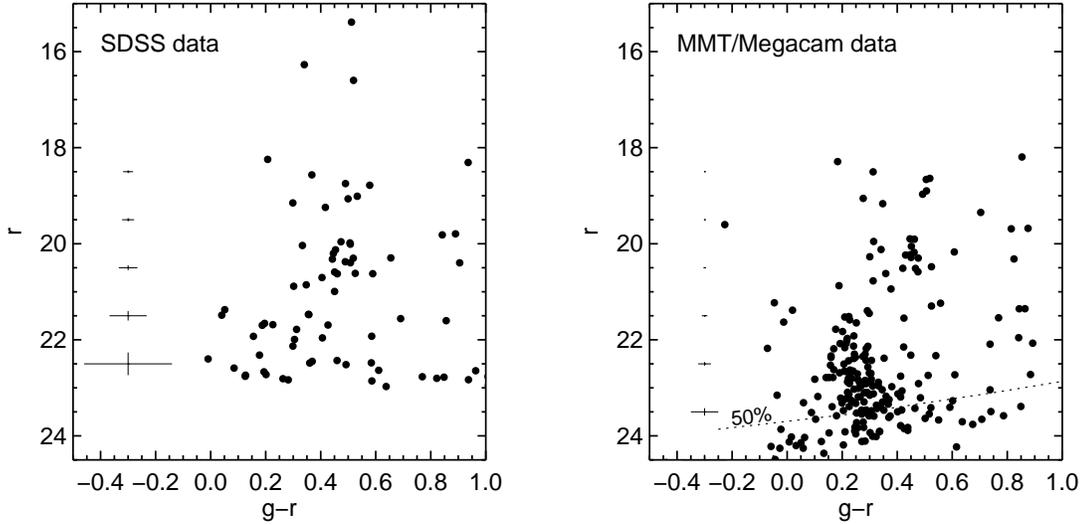} 
 \caption{Color-Magnitude Diagrams of stars within 4.5 arcmin  ($\sim1.5\times$ Plummer
  half-light radii) of the center of Bo\"otes
  II based on MMT/Megacam observations. The left CMD shows data from
  the SDSS DR6 and the right panel shows the MMT/Megacam data.  Error
  bars showing the color and magnitude uncertainties as a function of
  $r$ are overplotted.}
\label{fig:cmd}
\end{figure*}

Figure~\ref{fig:cmd} shows the CMDs of stars within 4.4 arcminutes of
the center of Bo\"otes II from the SDSS DR6 data and from the
MMT/Megacam data.  The 50\% completeness, as a function of color, is
overplotted on the Megacam CMD.

\section{Bo\"otes II Properties}

\subsection{Bootstrap Analysis}\label{bs}

With only $\sim$100 object stars resolved in this study, small number
statistics will constitute a substantial, if not dominant, source of
uncertainty in the derived quantities. For the ambiguous ultra-faint
satellites such as Bo\"otes II, a rigorous examination of the
uncertainties is essential to measure any of their properties. We use
a 10,000 iteration bootstrap analysis to determine both the most
likely values of Bo\"otes II's properties, and the uncertainties
associated with each measurement as detailed in the following
sections. For each iteration, the data are randomly re-sampled with
replacement and then analyzed to derive as described in the remainder
of \S3: central RA and Dec, distance modulus, Plummer and exponential
half-light radii and total absolute magnitudes, King core and tidal
radius, position angle, ellipticity, and asymmetry. Aside from King
tidal radius, the bootstrapped distributions of the derived parameters
are well described by a Gaussian. All quoted values and uncertainties
are thus the peaks and standard deviations of the bootstrap
distributions.  For the King tidal radius, we quote the half-width,
half-max as the uncertainty because there is a long, poorly populated
tail of values that extends to high tidal radius.  We report the
values of bootstrapped sample fits in Table 1.

\subsection{Central Position}

Bo\"otes II contains no detectable unresolved luminous component, so
we determine its center by locating the barycenter of likely member
stars within a 4.5 arcmin radius around the center published in Walsh
et al. (2007). We define likely member stars as those that have colors
and magnitudes consistent with M92 at an initial estimated distance
modulus. We repeat this with the returned RA and Dec until the
difference between input and output values converges. Our derived
values are presented in Table 1 along with their associated bootstrap
uncertainty.  The uncertainties on the center of Bo\"otes II are
substantial: 7.2 seconds of RA and 23 arcseconds in Dec.  Because all
parameters are derived for each of the bootstrapped samples, these
uncertainties on the center are automatically propagated through to
the uncertainties in Bo\"otes II's structural parameters.

\subsection{Distance} 

To investigate the distance to Bo\"otes II's stellar population, we
first compare its CMD to empirical globular cluster fiducials (M92,
M3, M13, and M71) with $-2.4 < [Fe/H] < -0.7$.  We use $m-M$ = 14.60,
15.14, 14.42 and 13.71 for the four clusters
\citep{paust07,kraft,cho,grundahl02}.  We choose to rely on fiducials,
rather than theoretical isochrones, because these well studied
globular clusters have photometry in the exact photometric system we
have calibrated our data to. The fiducials we use are based on those
of \citet{clem07} in Sloan $g' - r'$. They were converted into $g - r$
using the transformation of \citet{rider04} and checked by comparing
the transformed fiducials directly to the SDSS imaging of of the
clusters in the SDSS DR6 (J. Strader, private communication). The
robustness of our comparison depends on Bo\"otes II having an old
stellar population, like those in these four comparison clusters.  We
address and confirm this with theoretical isochrones in \S3.4.

For each fiducial, we find the distance modulus that provides the best
fit to the stars in the CMD shown in the right panel of
Figure~\ref{fig:cmd}.  This CMD includes all stars within 4.5 arcmin
of Bo\"otes II.  To determine this distance modulus, we step each
fiducial through 0.05 magnitude intervals in ($m-M$) from 17.5 to 19.5
mag and find the number of stars brighter than $r=23.5$ that,
considering color uncertainties, have colors within 0.05 magnitudes of
the fiducial. To eliminate the contribution of stars belonging to the
thick disk and halo, we then do the same for stars exterior to 9.0
arcmin and subtract this value normalized to an area of
$\pi(4.5')^2$. We take the best fit for each fiducial as the distance
modulus that maximizes this number of stars. The best fit distance
moduli for the M92, M3, M13 and M71 fiducials are 18.1, 18.1, 18.1 and
18.85 with 96, 77, 87 and 43 stars respectively.

\begin{figure*}[t!]
\plotone{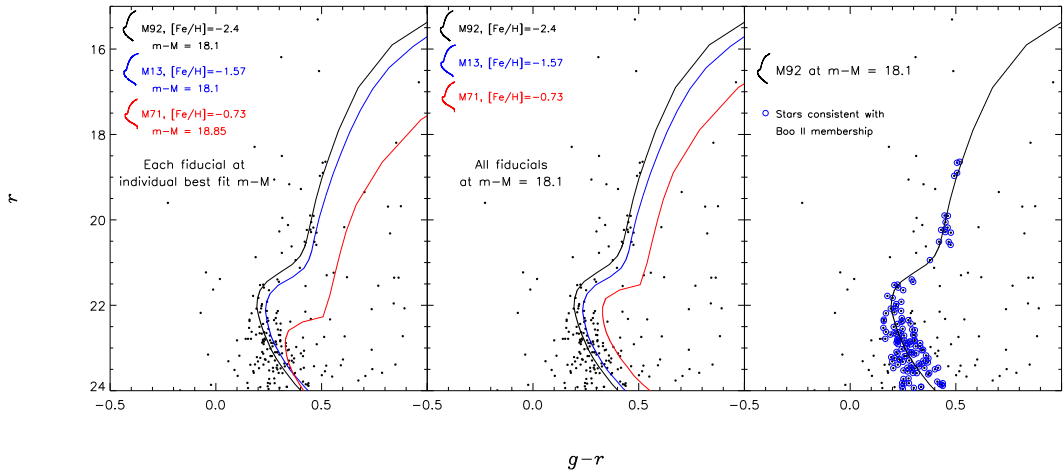} 
\caption{Color-Magnitude Diagrams of stars within 4.5 arcmin of the
  center of Bo\"otes II {\it Left}: Globular cluster fiducials
  overplotted at their own best fit distance modulus. {\it Center}: GC
  fiducials overplotted at M92 best fit distance modulus of $m-M=18.1$
  {\it Right}: M92 fiducial at $m-M=18.1$ with probable Bo\"otes II
  member stars highlighted.}
\label{fig:fancycmd}
\end{figure*}

The left panel of Figure~\ref{fig:fancycmd} shows the M92, M13 and M71
fiducials overplotted on the MMT CMD at their individual best fit
distances. For simplicity we include M13 and not M3, because they have
similar [Fe/H] and M13 provides a better match both quantitatively and
qualitatively. In the central panel of Figure \ref{fig:fancycmd} we
overplot the empirical M92, M13 and M71 fiducials, all projected to
the M92 best fit distance modulus of m-M = 18.1 (42 kpc).  These confirm
that it is reasonable to infer that the stellar population of Bo\"otes
II is like that of M92.  We therefore use the M92 fiducial in our
bootstrap analysis to derive the best-fit distance, which yields a
distance modulus of $m-M=18.1\pm 0.06$, only including the formal
bootstrap uncertainty, stemming from small number statistics. We add
in quadrature the distance modulus uncertainty of M92 (0.09
mag, \citealt{paust07}) and the $r$ zero-point uncertainty (0.025 mag, \S2)
to derive $m-M=18.1\pm 0.06$, or $d=42\pm2$ kpc. If Bo\"otes II has a
stellar population different from M92, then the uncertainty in the
distance is larger.

\subsection{Stellar Population}
M92 has a very low [Fe/H] of -2.4 and is $\alpha$-element enhanced
relative to solar ([Ca/Fe] = 0.3, \citealt{sneden00}). The match
between M92 and Bo\"otes II's stellar populations thus supports
[Fe/H]$_{BooII} \lta -2$, even if Bo\"otes II is $\alpha$-depleted
relative to M92 (typical of the contrast between dSph and globular
cluster populations, \citealt{alpha}).  Figure~\ref{fig:fancycmd}
shows that M13's fiducial sequence is only 0.04 mag redder in $(g-r)$
than that of M92 below the main-sequence turnoff.  The uncertainties
in the $g$ and $r$ zero-points of the MegaCam data in
Figure~\ref{fig:fancycmd} result in a $(g-r)$ calibration that is
uncertain at 0.038 mag.  These Bo\"otes II data are thus also
consistent with having a more moderate abundance ($[Fe/H] \sim
-1.6$). However, using independent $V$ and $I$ observations obtained
on VLT/FORS2, Jerjen et al. (in preparation) also find that Bo\"otes
II is best described by the most metal-poor fiducials.

To check these empirical results and to investigate a range of
possible ages for Bo\"otes II, we repeat the distance modulus fitting
described in \S3.3 using the theoretical isochrones of
\citet{dotter08} in SDSS colors. We use 24 isochrones corresponding to
the combinations with [Fe/H]$=-2.3$, $-1.5$ and $-0.7$, ages of 5, 7,
10 and 13 Gyr and alpha-element abundances of [$\alpha$/Fe] = 0.0 and
0.2. The best fitting of all 24 isochrones is that with [Fe/H]$=-2.3$,
13 Gyr and [$\alpha$/Fe] = 0.2, with 98 stars having colors lying
within 0.05 mags of the isochrone. For the same abundance values, the
number of stars drops to 86, 78 and 72 for the 10, 7 and 5 Gyr
populations respectively. This quantitative comparison with the Dotter
isochrones highlights the fact that the small color difference between
the MSTO and RGB stars in Bo\"otes II would not be consistent with a
stellar population much younger than 13 Gyr.

\subsection{Structural Parameters}

\begin{figure}
\plotone{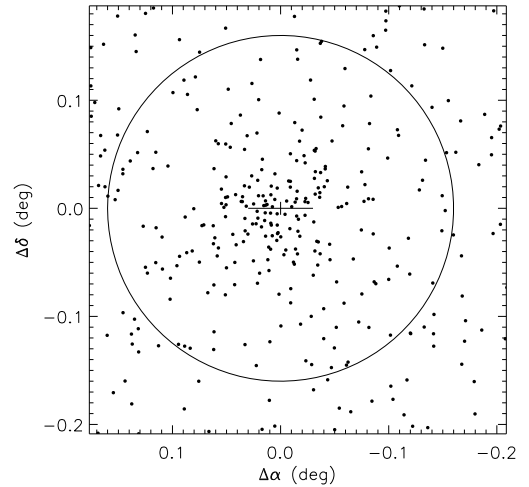} 
\caption{Positions of stars passing selection cuts. The calculated
center of Bo\"otes II is highlighted by the crosshair, which spans the RA and Dec uncertainties. The large circle shows the maximum radius of the surface density profile fit.
\label{fig:pos}}
\end{figure}

\begin{figure}
\plotone{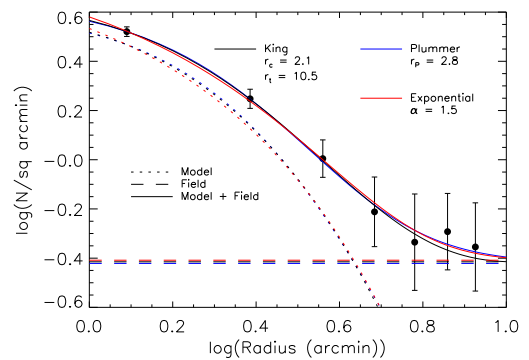} 
\caption{Fitted surface density profile of Bo\"otes II consisting
of a Plummer (blue), Exponential (red) or King (black) profile combined with a constant
field contribution. \label{fig:sbp}}
\end{figure}

The surface density profiles of globular clusters and dwarf spheroidal
galaxies (dSphs) are commonly parameterized by King \citep{king66},
Plummer \citep{plummer11} and exponential profiles.  To facilitate
comparison with other observational studies, we fit all three
profiles to the stellar distribution of Bo\"otes II:

\begin{equation}
\Sigma_{King}(r) = \Sigma_{0,K}\left( \left(1+\frac{r^2}{r_c^2}\right)^{-\frac{1}{2}}-\left(1+\frac{r_t^2}{r_c^2}\right)^{-\frac{1}{2}}\right)^2
\end{equation}

\begin{equation}
\Sigma_{Plummer}(r) =  \Sigma_{0,P}\left(1+\frac{r^2}{r_P^2}\right)^{-{2}}
\end{equation}

\begin{equation}
\Sigma_{exp}(r) =  \Sigma_{0,E}\exp\left(-\frac{r}{\alpha}\right)
\end{equation}
where $r_P$ and $\alpha$ are the scalelengths for the Plummer and
exponential profiles and $r_c$ and $r_t$ are the King core and tidal
radii, respectively. For the Plummer profile, $r_P$ equals the
half-light radius $r_h$, while for the exponential profile $r_h
\approx 1.668\alpha$.  The circled stars in the right panel of
Figure~\ref{fig:fancycmd} show the color-magnitude criteria we use to
select probable Bo\"otes II member stars for calculating its center
and for investigating its structure. Figure~\ref{fig:pos} shows the
spatial distribution of stars that pass these cuts, and the location
of their derived center. Figure~\ref{fig:sbp} shows the surface
density profile of Bo\"otes II around this center, where the error
bars were derived assuming Poisson statistics.  Using a non-linear
least squares method, we fit Plummer and exponential models plus a
constant field contamination to this surface density profile.  The
surface density profile fits are only constrained to be physically possible
systems (i.e. field contributions must be positive). In the case of
the King profile, the tidal radius is present in a constant term, hence
there is a degeneracy between the field value and the tidal radius. To
circumvent this we first fix the King field value using the mean
of the fitted Plummer and exponential field values.

All three fits yield consistent characteristic radii; Assuming a
distance modulus of $m-M=18.1$, the Plummer and exponential profiles
yield physical half-light radii of $r_{h, Plummer}\simeq36\pm 9$ pc and
$r_{h, exponential}\simeq33\pm 10$ pc. The King model fit yields a core
radius of $r_c\simeq25\pm 9$ pc and a tidal radius of $r_t\simeq155\pm
35$ pc. Although this tidal radius lies just outside the extent of the
radial profile we can measure, we find that the inner radial bins
constrain the core radius and central density while fixing the field
value leaves only the tidal radius as a free parameter. In the event
that the outer radial bins are contaminated with Boo II stars, and
therefore higher than the true field value, the tidal radius could be
much larger. As an example, if the true field value is an overestimate
of 25\% by the Plummer and exponential models, then the best fit tidal
radius is $\sim215$ pc while the core radius remains relatively
constant at $\sim22$pc. The fitted King tidal radius of $r_t\simeq155\pm
35$ is a lower limit.

\begin{figure}
\plotone{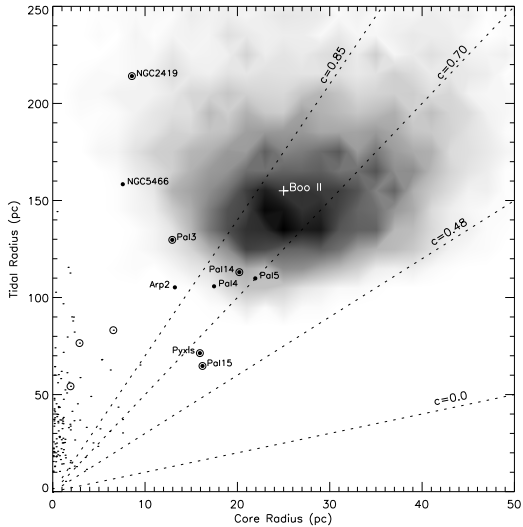} 
  \caption{2D Histogram of bootstrap results for Bo\"otes II's King
    core radius and tidal radius with Galactic globular clusters
    (GGCs).  GGCs having $r_c > 10 pc$ or $r_t > 100 pc$ are shown
    with larger points. Dashed lines show contours of constant
    concentration, defined as $c=\log(r_t/r_c)$.  The open circles
    show the 8 globulars with $35 < d < 100$ kpc for which we
    calculate M/L in \S4.3.  \label{fig:king}}
\end{figure}

Figure~\ref{fig:king} shows in greyscale the distribution of King core
and tidal radii derived from the bootstrap of Bo\"otes II stars.
Overplotted are the measured core and tidal radii of Galactic globular
clusters.  We have overplotted lines of constant concentration
($r_{tidal}/r_{core}$) and the concentrations of known globular
clusters calculated with the tidal and core radii in the catalog of
\citet{harris96}.  Only a handful of known globulars have $r_h\geq 10$
pc and the only one larger than $r_h\sim 20$ pc is known to be tidally
disrupting (Pal 5).  However, this figure shows several known globular
clusters with a King concentration as low as Bo\"otes II.  The King
concentrations of the Milky Way's dwarf spheroidals range from 0.48 --
1.12 \citep{mateo}, similar to the range of GCs in this figure. We do
not overplot the King parameters of the classical Milky Way dwarf
galaxies, because their relaxation times are $\sim$ a Hubble time.
Their King profile fits thus contain less physical meaning than those
of objects with shorter relaxation times, such as globular clusters
and Bo\"otes II (see \S4 for discussion).

\subsection{Luminosity}

To estimate the total luminosity of Bo\"otes II we integrate the model
components of the surface density profile, corrected for
incompleteness, to derive the number of Bo\"otes II stars within the
CMD selection cuts. Using this to normalize the theoretical luminosity
function gives estimates of $M_r=-2.6$ ($M_V=-2.4$) for the Plummer
profile and $M_r=-2.4$ ($M_V=-2.2$) for the exponential profile (using
$V-r=0.16$, adapted from \citealt{sdssiso} for a 13 Gyr, [Fe/H] =
-2.27 stellar population) . We obtain these magnitudes after
correcting for the missing flux from stars fainter than $r=23.5$ by
integrating the theoretical luminosity function taken from
\cite{sdssiso} for a [Fe/H]$=-2.27$, 13 Gyr population. This method
counts the number of stars without regard to their individual
magnitudes, which in systems of such low luminosity could strongly be
affected by the addition or subtraction of a single RGB star. Such a
star's individual magnitude could be much brighter than $M_V=-2$,
rendering a more traditional summing of fluxes method unreliable. As
the total luminosity of a stellar system becomes comparable to the
luminosity of individual stars, the summed luminosity becomes
dominated by the brightest stars which may or may not be present
simply from small number statistics. Hence, a group of stellar systems
with an identical number of stars in each will have a spread in
luminosity independent of any measurement uncertainty (see also
\citet{martin08}). Using simulated dwarf galaxies from Walsh, Willman
\& Jerjen (\emph{in prep}) we find that for an object such as Bo\"otes
II as observed by MMT/Megacam, this spread has a standard deviation of
0.6 magnitudes. We therefore combine this effect with the bootstrap
uncertainty by summing in quadrature to derive the values presented in
Table 1.

\subsection{Morphology}

\begin{figure}
\plotone{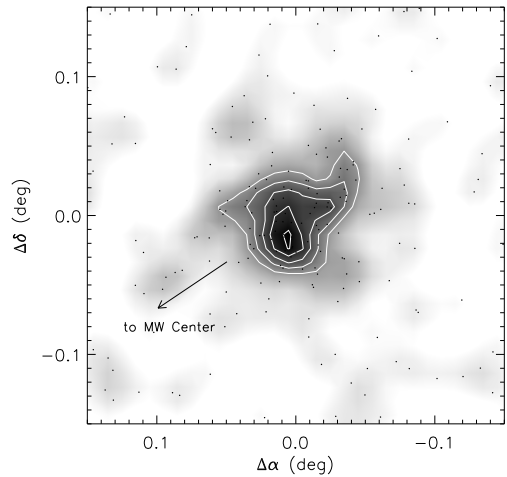} 
\caption{Smoothed contour plot of Bo\"otes II showing an apparent
distortion along the direction towards the Galactic Center. The
contours show the 3,4,5 and 6 sigma levels above the mean.
  \label{fig:cont}}
\end{figure}

We look for evidence of tidal disturbance based on the morphology of
Bo\"otes II's isodensity contours.  Binning the positions of stars in
Figure~\ref{fig:pos} into $0.01^\circ\times0.01^\circ$ bins and
spatially smoothing with a Gaussian of 1.5 arcmin FWHM scale length
reveals an apparent distorted morphology to Bo\"otes II, including
substructure at the 3 - 5 $\sigma$ level and an elongation directed
along the gradient of the Galactic potential, as shown in
Figure~\ref{fig:cont}.  Although tidal tails trace an object's orbit,
tidal debris is stripped from an object along the gradient of the
gravitational potential such that tidal stars near an object are
expected to lie along this gradient.  The elongation of Bo\"otes II
could thus be a feature resulting from tidal interaction.

\begin{figure*}[t!]
%\epsscale{0.8}
\plotone{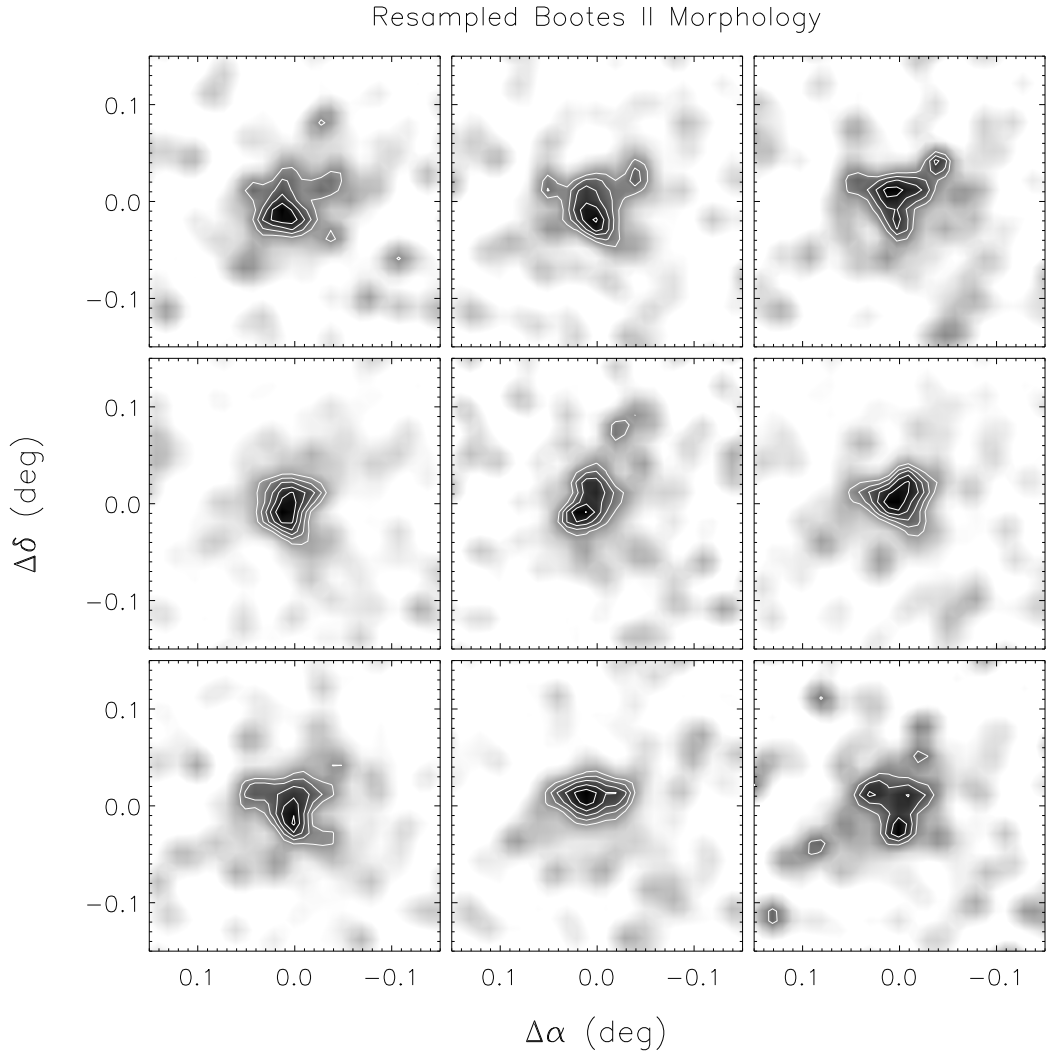} 
%\plotone{boosim3.ps} 
 \caption{Smoothed contour plot of 9 randomly selected resamples of
  Bo\"otes II stars, showing varying morphology for each iteration.
  The contours show the 3,4,5,6 and 7 sigma levels above the mean.
  \label{fig:sim}}
\end{figure*}

However, due to the meager number of stars in Boo II, any observed
irregular morphology may be an effect of small number statistics.  To
evaluate the significance of the morphology shown in
Figure~\ref{fig:cont}, we generate contour plots of bootstrap
resamples of Bo\"otes II stars.  Figure \ref{fig:sim} shows nine such
randomly selected isodensity contours defined the same way as in
Figure~\ref{fig:cont}. This figure shows that the irregular morphology
and apparent distortion along the Galactic potential are not
persistent features. While this does not rule out tidal disturbance to
Bo\"otes II, the varying morphologies of these resampled objects
demonstrates that the shape of Bo\"otes II does not necessarily
reflect a true irregularity in its underlying spatial distribution.

In order to quantify any asymmetry in the morphology of Bo\"otes II, we
first derive the position angle $\theta$ and ellipticity $e$. We
calculate these from the standard SExtractor definitions
\citep{bertin96}, using the smoothed images assuming that all pixels
greater than $3\sigma$ above the mean are part of Bo\"otes II. We then
count the number of stars on either side of the major and minor axes,
within 1.5$r_h$. If the positions of these stars are drawn from an
axisymmetric distribution, then the numbers on either side of an axis
should be within $\sqrt{2 \langle N\rangle}$ of each other. We
define an asymmetry parameter $A$:
\begin{equation}
A = \frac{N_1 - N_2}{\sqrt{2 \langle N\rangle}}
\end{equation}
where $N_1$ and $N_2$ are the counts on either side of the axis and
$\langle N\rangle$ is their average. Hence, for each bootstrap
iteration we have two values of $A$, one for
the major and one for the minor axis. Doing this for simulated Boo
II-like objects drawn from a pure Plummer density profile expectedly
yields a distribution of $A$ with a mean of 0.0 and a standard
deviation of $\sigma_{sims} = 1.01$. The bootstrap yields two
distributions with means of $A_{minor}=-0.23$ and $A_{major}=-0.18$
and standard deviations of $\sigma_{minor}= 1.03$ and
$\sigma_{major}=1.01$. The asymmetry of Bo\"otes II is therefore not
statistically significant.

These results imply that any apparent asymmetry in the distribution of
Bo\"otes II stars is well within that expected from a symmetric
system, and is probably due to small number statistics. Deeper and
wider-field imaging may provide the signal necessary to definitively
measure whether Bo\"otes II has extended, tidal structure.

\section{Is Bo\"otes II dark matter dominated?}

As discussed in the Introduction, one motivation for studying an
individual ultra-low luminosity object such as Bo\"otes II in great
detail is to determine whether it is a galaxy or a star cluster. Using
our definition of a dwarf spheroidal galaxy, this boils down to
determining whether or not the object is dark-matter dominated.
(However, for example, see \citealt{metz07,dabringhausen08} for
another interpretation of the Milky Way's dwarf companions).
Understanding the properties of dark matter depends critically on
identifying the smallest mass and length scales at which dark matter
clusters
\citep[e.g.][]{strigari07c,gilmore07}. The most direct way we have at
present to investigate dark matter on the smallest scales is by using
the least luminous galaxies as tracers of dark matter, and by pushing
the envelope to find the smallest galaxies possible to form. 

In this section, we use a tidal argument to show why Bo\"otes II may
be dark matter dominated, and thus a dwarf galaxy, despite its very
low luminosity ($L \sim 500 L_{\odot}$).  The lines of evidence
presented in this section are circumstantial now, but are illustrative
of arguments that could provide strong constraints on the correct
classification for Bo\"otes II, if deeper, wider-field imaging,
kinematics, and/or spectroscopic abundance measurements are able to
demonstrate whether Bo\"otes II is self-bound or not.

\begin{figure}
\plotone{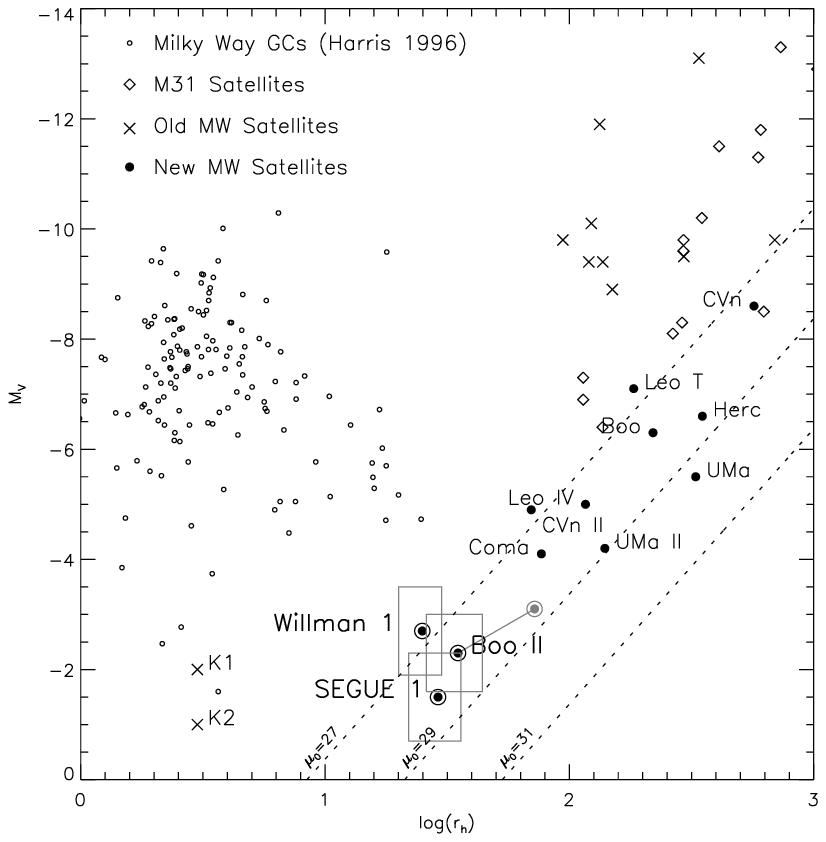} 
\caption{Size-luminosity plot of known Galactic satellites. The
  previous position of Bo\"otes II is shown in grey along with the
  region bounded by the uncertainties in $r_h$ and $M_V$. Values for
  the new Milky Way satellites are taken from
  \citet{martin08}. Globular cluster properties are from Harris
  (1996). Dotted lines show lines of constant surface
  brightness.  \label{fig:lumrad}}
\end{figure}

\subsection{Separation in size-luminosity space from dwarf galaxies and globular clusters}

In Figure~\ref{fig:lumrad}, we compare the size and luminosity of
Bo\"otes II with that of other nearby, old, stellar populations.  This
figure shows Bo\"otes II in the size-luminosity plane along with Milky
Way globular clusters \citep[GCs]{harris96,k12} and M31
\citep{andprops,and9,and10} and Milky Way dwarf satellites
\citep{mateo,grebelLG,quintet,uma,uma2,willman1,leot,cvn,bootes}. In
this figure, including only known old stellar populations, Bo\"otes II,
Wil1 and Segue 1 occupy a somewhat unique place in size-luminosity
space.  The apparent lack of objects with half-light radii larger than
those of Bo\"otes II, W1 and Segue 1 but smaller than the apparent
minimum size of confirmed dwarf galaxies ($\sim$ 100 pc) is quite
possibly an observational selection effect due to the faint central
surface brightnesses of objects in that region of the size-luminosity
plane.

Conversely, Bo\"otes II, Wil1, and Segue 1 have half-light radii (20
-- 40 pc) an order of magnitude larger than the half-light size
characteristic of similarly low luminosity globular clusters ($-4 <
M_V < -1$.) Low luminosity Milky Way clusters in that size gap could
have been detected by the SDSS searches of \cite{koplf} or Walsh,
Willman \& Jerjen (\emph{in prep}).  This gap between Bo\"otes II,
Willman 1 and Segue 1 and the known globular clusters is thus real and
not a selection effect, suggesting that these objects are a distinct
population from Milky Way GCs.  However, if such diffuse old clusters
do exist outside of the Milky Way, they would not yet have been
discovered owing to their low surface brightnesses.  It thus remains
possible that old star clusters with properties bridging the gap
between Bo\"otes II and known GCs do exist in abundance in other
environments.

Based on the compilation of \citet{dias02} (catalog obtained at
http://www.astro.iag.usp.br/~wilton/), only 2 of the 1076 Milky Way
open clusters in that catalog with both distance and diameter
measurements have apparent radii larger than 20 pc.  Young stellar
associations in environments other than the Milky Way have been
observed to have characteristic sizes of up to 100 pc (e.g. OB
associations in the LMC observed by \citealt{gouliermis03}).  Although
the majority of these clusters appear to be unbound, some fraction of
them could survive as self-bound entities for 10 Gyr if subject only
to very weak tidal forces. We will show in \S4.2 and 4.3, that tidal
effects limit the survivability of objects at Bo\"otes II distances
with star cluster-like mass-to-light ratios and sizes and luminosities
similar to those of Bo\"otes II, Willman 1, and Segue 1.

\subsection{Minimum mass-to-light ratio to be bound}

We estimate a lower limit to the mass-to-light ratio required for
Bo\"otes II to be self-bound. We calculate the instantaneous tidal
radius of Bo\"otes II using:
\begin{equation}
r_{tidal} \varpropto \left(\frac{m_{BooII}}{3M_{MW}(d)}\right)^\frac{1}{3}d_{BooII},
\end{equation}
where $m_{BooII}$ is the mass of Bo\"otes II, $M_{MW}(d)$ is the mass
of the Milky Way within the distance to Bo\"otes II, and $d_{BooII}$ is
the galactocentric distance to Bo\"otes II.  We calculate the Milky Way mass
$M_{MW}(d)$ assuming an isothermal sphere with a circular velocity of
$V_c=220\pm40$ km s$^{-1}$ \citep{mwmass}, using $d_{BooII} = 42\pm2$ kpc,
and using $M_V = -2.4\pm0.7$ mag for Bo\"otes II .  For $(M/L)_V$ values of
2, 10, 100 and 1000\,kpc the corresponding tidal radii are thus
42$^{+19}_{-13}$, 72$^{+33}_{-21}$, 156$^{+70}_{-46}$ and
336$^{+151}_{-99}$ pc respectively. A tidal radius of 72 pc ($M/L=10$)
is within our observed field and only $\sim2$ times the half-light
radius of Bo\"otes II.  The visible extent of Bo\"otes II thus exceeds
that expected for globular cluster-like mass-to-light ratios.  If the
Bo\"otes II stars throughout our field-of-view are indeed bound to the
object, then it requires a significant dark matter component.  In
\S3.7, we showed that Bo\"otes II lacks statistically significant
tidal distortion, so there is presently no evidence that is is losing
stars (however see \citealt{munoz08}). Deeper wide-field imaging and/or
kinematic data will provide a clearer picture of the boundedness of
this object.

\subsection{Mass-to-light ratio inferred from the King tidal radius} 

Using the same physical principle as in the previous section, we now
calculate the mass-to-light ratio of Bo\"otes II assuming that the
King tidal radius derived in \S3.5 is the true tidal radius of Bo\"otes
II, owing to the tidal field of the Milky Way.  The King model
\citep{king66} is physically motivated and expected for a relaxed,
single mass component, spherical system in equilibrium that is tidally
limited by the gravitational field of the Milky Way.  For velocity
dispersions greater than 0.1 km/sec, the relaxation time of Bo\"otes
II is less than a Hubble time (unlike the classical Milky Way
dSphs). Based on a \citet{dotter08} stellar luminosity function for a
13 Gyr, $[Fe/H] = -2$ system normalized to have $M_V$ = -2.4 mag,
Bo\"otes II has 3700 stars more massive than 0.1$M_{Sun}$.  Its
relaxation time ($t_{relax} \sim N/8ln(N) * t_{cross}$;
\citealt{binney87}) at its Plummer half-light radius of 36 pc is thus
$\sim$ 1.1 Gyr/$\sigma_v$, where $\sigma$ is its 1D velocity
dispersion.  Although the presence of a significant dark component of
matter would complicate the expected relaxation time of Bo\"otes II,
we are testing the hypothesis that it does not contain dark matter.

If we take the King ``tidal radius'' to be the instantaneous tidal
radius of Bo\"otes II, we can infer a mass $m_{BooII}$ and therefore a
mass-to-light ratio.  We use Equation 7 as described in \S4.2, but set
$r_{tidal}=r_{tidal,King}=155\pm35$ pc, and solve for the mass of
Bo\"otes II. We infer a V-band mass-to-light ratio of Bo\"otes II of
$98^{+420}_{-84}$. If this is representative of the true $M/L$ of
Bo\"otes II, its mass would be dominated by dark matter, classifying
it as a galaxy.  We note the substantial uncertainty on the $M/L$
derived in this way.  Even if the input assumptions are robust, the
dark matter signal is only significant at the 1-$\sigma$ level.  As
mentioned in \S3.5, the King tidal radius may be substantially larger
if the outer surface density bins are contaminated with Boo II stars
thereby overestimating the field density. As in \S3.5, if the field
value is overestimated by 25\%, then the best fit tidal radius would
be $r_t=215$ pc and the inferred $M/L$ would be $263^{+991}_{-216}$.

% from \S4.2 and assume a
%circular orbit at Galactocentric radius of $d=42\pm2$ kpc,
%$V_c=220\pm40$ kms$^{-1}$,

For comparison, we use these same assumptions and same technique to
calculate the mass-to-light ratios of known Milky Way globular
clusters in the distance range of $35\leq R_{GC} \leq 100$ kpc.  We
chose this distance range because i) these halo globulars reside in an
environment and are on orbits that are the most comparable to Bo\"otes
II, and ii) \cite{mwmass} states the Milky Way mass profile is
consistent with an isothermal sphere with a circular velocity of
$V_c=220\pm40$ kms$^{-1}$ between 35 -- 100 kpc.  There are 8 clusters
in that range: Eridanus, Pal 2, NGC 2419, Pyxis, Pal 3, Pal 14, Pal 15
and NGC 7006. For all of these clusters, we use the distances and
luminosities from the \citet{harris96} catalog, and assign a distance
uncertainty of 0.1 mag in distance modulus. For Pal 2, NGC 2419, Pal
3, Pal 14 and NGC 7006 we use the core radii, concentrations and
uncertainties from \citet{mclaughlin05}. For the other three GCs, we
use values from \citet{harris96} and assign uncertainties of 10 pc to
their tidal radii. We find a median (mean) $(M/L)_V$ of
0.36$^{+0.98}_{-0.12}$ (0.5$^{+1.85}_{-0.38}$) for these eight
clusters. The single outlier with a calculated $M/L > 1$ is Pal 14
with $M/L=2.4^{+10.4}_{-1.9}$.  Although the true $M/L$ of most (all)
of these eight halo globulars is within 1$\sigma$ (2$\sigma$) of the
calculated values, we find that this technique systematically
underestimates of the $M/L$ of these relaxed systems by up to an order
of magnitude (3 of the eight GCs have inferred $M/L < 0.1$).  This
systematic underestimate is not surprising, because the tidal forces
experienced by these halo globulars at this snapshot in time are
smaller than the maximum tidal force that they have experienced in
their past.  Using their present distance in this $M/L$ calculation
will thus provide a lower limit on the masses required to yield the
observed King tidal radii.  Regardless, this calculation shows that
Bo\"otes II is an outlier from globular clusters with this metric of
measurement.

\section{Conclusion}

In this paper, we use MMT/MegaCam imaging in $g$ and $r$ to present
the first robust estimates of the fundamental properties of the
ultra-low luminosity Milky Way satellite Bo\"otes II (d $\sim$ 42
kpc).  This object is old and its stellar population appears very
similar to that of M92, showing that it is metal-poor ([Fe/H] $\lta
-2$).  With a total luminosity of only $\sim 500$ solar luminosities
($M_V \sim -2.4 \pm 0.7$ mag) and a half-light size of $\sim 36\pm9$
pc (assuming a Plummer profile), Bo\"otes II lies away from globular
clusters and dwarf spheroidals, but near Willman 1 and Segue 1, in
size-luminosity space. We showed that although the morphology of
Bo\"otes II appears irregular and elongated along the direction of the
Galactic potential, that this distortion of its isodensity contours is
not statistically significant in our dataset.

The revised values we present for the distance, luminosity, and
physical size of Bo\"otes II differ from those originally estimated in
the \citet{boo2} discovery paper. The primary factor in these
differences is the new distance estimate is 42 kpc rather than 60 kpc.
The SDSS discovery data was more than two magnitudes shallower than
the data presented in this paper.  VLT photometry of Bo\"otes II will
be presented in Jerjen et al. (\emph{in prep.}) along with a more
detailed discussion of its stellar population and of the possible
association, or lack thereof, with the Sagittarius Stream

Our bootstrap analysis demonstrated the impact of small number
statistics on the derived parameters for this ultra-faint class of
objects, but showed that despite large uncertainties Bo\"otes II has a
size that makes it distinct from Milky Way GCs, although its King
concentration is similar to that of Milky Way dwarf galaxies as well
as to some diffuse GCs.

The gap between Bo\"otes II, Willman 1, and Segue 1 and Milky Way
  globular clusters in size-luminosity space is not a selection effect
  because existing surveys would have been sensitive to such objects.
  However, the apparent separation in half-light size at 100 pc
  between dSphs and Bo\"otes II, Willman 1, and Segue 1 could be a
  selection effect, making it more likely that these three objects are
  fundamentally connected with the dwarf galaxy population. We pointed
  out that old star clusters filling in that apparent gap may exist in
  environments other than the Milky Way, but would have escaped
  detection owing to their low surface brightnesses.

If Bo\"otes II is a self-bound system in equilibrium, it could
represent the continuation of the dwarf galaxy population into the
extreme low luminosity regime. We showed that it is reasonable to
believe that Bo\"otes II is a relaxed system, and used its King tidal
radius to infer a lower limit M/L of $\sim98^{+420}_{-84}$. Dropping
the assumption that the King tidal radius is physically meaningful for
Bo\"otes II, its spatial extent is larger than that naively expected
for globular cluster-like mass-to-light ratios, if it is self-bound.
However, we cannot rule out that it is a low $M/L$ star cluster that
is undergoing tidal disruption or disturbance, much like the Pal 5
cluster.

Regardless of whether or not Bo\"otes II is dark matter dominated, it
can provide a unique laboratory with which to investigate the low
surface density limit of star formation and the tidal field of our
Galaxy.  The study in this paper also provides a partial road map for
the future study of numerous similar objects that may be discovered in
upcoming surveys for Galactic satellites to greater depth
(e.g. PanSTARRS, \citealt{panstarrs}) or in previously unsearched sky
(e.g. Skymapper, \citealt{skymapper} and Walsh et al. \emph{in
prep.}).  Ultimately, determining the boundedness of its stars will be
needed to definitively pin down the nature of the peculiar Bo\"otes II
object.

\acknowledgments

We thank the anonymous referee for improving this manuscript. We thank
Bill Wyatt for maintaining a local copy of the SDSS data products at
the Harvard-Smithsonian Center for Astrophysics.  Jay Strader shared
his transformations of the Clem (2007) isochrones used in this paper,
as well as provided helpful conversations. We thank Brian McLeod for
providing guidance and troubleshooting during our observing runs,
writing software that our reduction relied on, and advising our
MegaCam reduction.  We thank Matt Ashby for creating the MegaCam
reduction manual and for advising our reduction. We thank Maureen
Conroy for troubleshooting during our observing runs, and for helping
to create weight maps and performing a preliminary co-addition of our
Bootes II images.  Mike Alegria, John McAfee, and Ale Milone provided
observational support during our observing runs.  Observations
reported here were obtained at the MMT Observatory, a joint facility
of the Smithsonian Institution and the University of Arizona. SW
acknowledges partial financial support from the Australian Research
Council Discovery Project Grant DP0451426.

{\it Facilities:} \facility{MMT (MegaCam)}

%\bibliographystyle{../../apj}
%\bibliography{paper}

%\clearpage

\begin{deluxetable}{lccc}
\label{tab:info}
\tablecaption{Bo\"{o}tes II Properties} 
\tablewidth{0pt} 
\tablehead{ 
%\multicolumn{4}{c}{Bo\"{o}tes II Properties}
\colhead{Parameter} & \colhead{Measured} & \colhead{Uncertainty} & \colhead{bootstrap median}
}

\startdata 

RA (h m s)         &13 58 05.1         & $\pm7.2$s        & 13 58 04.3\\
Dec (d m s)        &+12 51 31        & $\pm23\prime\prime$    & +12 51 09\\
$(l,b)$         & (353.75,68.86)     &-            & -\\
$(m-M)$            & $18.1$         &$\pm0.06$        & 18.1\\
Distance         & $42$\,kpc         &$\pm1.6$\,kpc        & 42\,kpc\\
$r_{h}$ (Pl)         & $2.8$\,arcmin        &    $\pm0.7$    & 2.8\,arcmin\\
$r_{h}$ (ex)         & $2.5$\,arcmin        &    $\pm0.8$    & 2.6\,arcmin\\
$r_{h}$ (Pl)         & $35$\,pc         &    $\pm9$        & 36\,pc\\
$r_{h}$ (ex)         & $31$\,pc        &    $\pm10$        & 33\,pc\\
$M_{V}$ (Pl)         & $-2.3 $         &$\pm0.7$        & $-2.4 $\\
$M_{V}$ (ex)         & $-2.2 $         &    $\pm0.7$    & $-2.2 $\\
$\mu_{0,\rm V}$ (Pl)    & $27.76$\,mag arcsec$^{-2}$ & $\pm 0.31$     & $27.93$\,mag arcsec$^{-2}$\\
$\mu_{0,\rm V}$ (ex)    & $27.70$\,mag arcsec$^{-2}$ & $\pm 0.33$     & $27.90$\,mag arcsec$^{-2}$\\
$r_{c}$         & $25$\,pc         &$\pm9$            & 25\,pc\\
$r_{t}$         & $127$\,pc         &$\pm35$        & 155\,pc\\
$\theta$         & $-33^{\circ}$     &$\pm57^{\circ}$    & $-28.5^{\circ}$ \\
$e$             & $0.27$        &$\pm0.15$        & $0.34$    \\
$A_{major}$        & -            &    $\pm1.01$    & $-0.26$    \\
$A_{minor}$        & -            &    $\pm1.03$    & $-0.33$    \\

\enddata
\end{deluxetable}

\end{document}